\documentclass[aps,prl,twocolumn,superscriptaddress,nofootinbib]{revtex4}

\usepackage{graphicx}
\usepackage[breaklinks=true]{hyperref}
\usepackage{xcolor}

\begin{document}

\title{First direct measurement constraining the $^{34}$Ar($\alpha$,p)$^{37}$K reaction cross section for mixed hydrogen and helium burning in accreting neutron stars}

\author{J. Browne}
\affiliation{Department of Physics and Astronomy and National Superconducting Cyclotron Laboratory, Michigan State University, East Lansing, MI 48824}
\affiliation{Joint Institute for Nuclear Astrophysics (JINA-CEE)}
\author{K.A. Chipps}
\altaffiliation{Corresponding author}
\affiliation{Physics Division, Oak Ridge National Laboratory, Oak Ridge, TN 37831}
\affiliation{Department of Physics and Astronomy, University of Tennessee, Knoxville, TN 37996}
\author{Konrad Schmidt}
\affiliation{Department of Physics and Astronomy and National Superconducting Cyclotron Laboratory, Michigan State University, East Lansing, MI 48824}
\affiliation{Joint Institute for Nuclear Astrophysics (JINA-CEE)}
\affiliation{Institute of Radiation Physics, Helmholtz-Zentrum Dresden-Rossendorf, Bautzner Landstrasse 400, 01328 Dresden, Germany}
\author{H. Schatz}
\affiliation{Department of Physics and Astronomy and National Superconducting Cyclotron Laboratory, Michigan State University, East Lansing, MI 48824}
\affiliation{Joint Institute for Nuclear Astrophysics (JINA-CEE)}
\author{S. Ahn}
\altaffiliation{Current address: Center for Exotic Nuclear Studies, Institute for Basic Science, 34126 Daejeon, South Korea}
\affiliation{Department of Physics and Astronomy and National Superconducting Cyclotron Laboratory, Michigan State University, East Lansing, MI 48824}
\affiliation{Joint Institute for Nuclear Astrophysics (JINA-CEE)}
\author{S.D. Pain}
\affiliation{Physics Division, Oak Ridge National Laboratory, Oak Ridge, TN 37831}
\affiliation{Department of Physics and Astronomy, University of Tennessee, Knoxville, TN 37996}
\author{F. Montes}
\affiliation{Department of Physics and Astronomy and National Superconducting Cyclotron Laboratory, Michigan State University, East Lansing, MI 48824}
\affiliation{Joint Institute for Nuclear Astrophysics (JINA-CEE)}
\author{W.J. Ong}
\altaffiliation{Current address: Nuclear and Chemical Sciences Division, Lawrence Livermore National Laboratory, Livermore, CA 94550}
\affiliation{Department of Physics and Astronomy and National Superconducting Cyclotron Laboratory, Michigan State University, East Lansing, MI 48824}
\affiliation{Joint Institute for Nuclear Astrophysics (JINA-CEE)}
\author{U. Greife}
\affiliation{Physics Department, Colorado School of Mines, Golden, CO 80401}

\author{J. Allen}
\affiliation{Dept. of Physics and Astronomy, University of Notre Dame, Notre Dame, IN 46556}
\author{D.W. Bardayan}
\affiliation{Dept. of Physics and Astronomy, University of Notre Dame, Notre Dame, IN 46556}
\affiliation{Joint Institute for Nuclear Astrophysics (JINA-CEE)}
\author{J.C. Blackmon}
\affiliation{Department of Physics and Astronomy, Louisiana State University, Baton Rouge, LA, 70803}
\author{D. Blankstein}
\affiliation{Dept. of Physics and Astronomy, University of Notre Dame, Notre Dame, IN 46556}
\author{S. Cha}
\altaffiliation{Current address: Center for Exotic Nuclear Studies, Institute for Basic Science, 34126 Daejeon, South Korea}
\affiliation{Department of Physics, Sungkyunkwan University, Suwon 16419, Korea}
\author{K.Y. Chae}
\affiliation{Department of Physics, Sungkyunkwan University, Suwon 16419, Korea}
\author{M. Febbraro}
\affiliation{Physics Division, Oak Ridge National Laboratory, Oak Ridge, TN 37831}
\author{M.R. Hall}
\affiliation{Dept. of Physics and Astronomy, University of Notre Dame, Notre Dame, IN 46556}
\author{K.L. Jones}
\affiliation{Department of Physics and Astronomy, University of Tennessee, Knoxville, TN 37996}
\author{A. Kontos}
\affiliation{Department of Physics and Astronomy and National Superconducting Cyclotron Laboratory, Michigan State University, East Lansing, MI 48824}
\affiliation{Joint Institute for Nuclear Astrophysics (JINA-CEE)}
\author{Z. Meisel}
\affiliation{Department of Physics and Astronomy and National Superconducting Cyclotron Laboratory, Michigan State University, East Lansing, MI 48824}
\affiliation{Joint Institute for Nuclear Astrophysics (JINA-CEE)}
\author{P.D. O'Malley}
\affiliation{Dept. of Physics and Astronomy, University of Notre Dame, Notre Dame, IN 46556}
\author{K.T. Schmitt}
\affiliation{Physics Division, Oak Ridge National Laboratory, Oak Ridge, TN 37831}
\author{K. Smith}
\affiliation{Department of Physics and Astronomy, University of Tennessee, Knoxville, TN 37996}
\author{M.S. Smith}
\affiliation{Physics Division, Oak Ridge National Laboratory, Oak Ridge, TN 37831}
\author{P. Thompson}
\affiliation{Department of Physics and Astronomy, University of Tennessee, Knoxville, TN 37996}
\author{R. Toomey}
\affiliation{Department of Physics and Astronomy, Rutgers University, Piscataway, NJ, 08854}
\author{M. Vostinar}
\affiliation{Department of Physics and Astronomy, Rutgers University, Piscataway, NJ, 08854}
\author{D. Walter}
\affiliation{Department of Physics and Astronomy, Rutgers University, Piscataway, NJ, 08854}

\collaboration{JENSA Collaboration}
\noaffiliation

\date{\today}

\begin{abstract}
The rate of the final step in the astrophysical $\alpha$p-process, the $^{34}$Ar($\alpha$,\textit{p})$^{37}$K reaction, suffers from large uncertainties due to lack of experimental data, despite having a considerable impact on the observable light curves of x-ray bursts and the composition of the ashes of hydrogen and helium burning on accreting neutron stars.  We present the first direct measurement constraining the $^{34}$Ar($\alpha$,p)$^{37}$K reaction cross section, using the Jet Experiments in Nuclear Structure and Astrophysics (JENSA) gas jet target. The combined cross section for the $^{34}$Ar,Cl($\alpha$,p)$^{37}$K,Ar reaction is found to agree well with Hauser-Feshbach predictions. The $^{34}$Ar($\alpha$,2p)$^{36}$Ar cross section, which can be exclusively attributed to the $^{34}$Ar beam component, also agrees to within the typical uncertainties quoted for statistical models. This indicates the applicability of the statistical model for predicting astrophysical ($\alpha$,p) reaction rates in this part of the $\alpha$p process, in contrast to earlier findings from indirect reaction studies indicating orders-of-magnitude discrepancies. This removes a significant uncertainty in models of hydrogen and helium burning on accreting neutron stars.
\end{abstract}

\maketitle

\pagebreak


Mixed hydrogen and helium burning at extreme temperatures and densities occurs on the surface of accreting neutron stars in low mass x-ray binaries \cite{Schatz2006,Meisel2018,Galloway2021}. At lower accretion rates, the nuclear fuel from the companion star accumulates on the neutron star until ignition of nuclear reactions triggers a thermonuclear runaway. Explosive burning of the accreted layer powers frequently observed x-ray bursts with recurrence times of hours to days. At higher accretion rates, thermonuclear reactions burn hydrogen and helium stably in steady state. This occurs either in systems with high global accretion rates, possibly associated with x-ray binaries observed to be in a soft spectral state, or in x-ray pulsars where strong magnetic fields restrict the accretion flow to a smaller surface area. 

For both explosive and steady state burning, temperatures over $\sim$1 GK can be achieved, triggering reaction chains which break out of the hot-CNO process into the $\alpha$p-process. The main reaction sequence follows:\\
\\
$^{18}$Ne$(\alpha,p)^{21}$Na$(p,\gamma)^{22}$Mg$(\alpha,p)^{25}$Al$(p,\gamma)^{26}$Si$(\alpha,p)^{29}$P$\\(p,\gamma)^{30}$S$(\alpha,p)^{33}$Cl$(p,\gamma)^{34}$Ar$(\alpha,p)^{37}$K\\
\\
which then leads into the rapid proton capture process (rp-process) \cite{Wallace1981,Schatz1999,Schatz1999b,Fis08}. The speed and extent of the $\alpha$p process determine the amount of helium burning early in the burst, the helium burning time structure, and the hydrogen-to-seed-nucleus ratio for the rp-process \cite{Schatz1999}. The latter determines the heaviest elements reached by the rp-process and therefore the composition of the nuclear ashes and neutron star crust \cite{Schatz1999,Schatz2001,Gupta08}. It also directly affects the late time behavior of the burst light curve \cite{Schatz2001,Woosley2004,Jose2010}.

Nuclear physics uncertainties in the $\alpha$p-process must be reduced to compare x-ray burst models with observed light curves and predict the composition of the produced nuclear ashes. Light curve model comparisons can be used to extract parameters such as the accreted composition or neutron star properties \cite{Zand2017,Meisel2018}. The composition of the produced nuclear ashes is needed for predictions of nuclear heating and cooling and thermal conductivity in the neutron star crust \cite{Gupta08,Sch14}, and to interpret observations of cooling neutron stars in transiently accreting systems. Predictions for composition of the ashes from steady state burning are needed to complement burst contributions in systems with episodes of stable burning. They are also relevant for constraining thermal and electrical conductivities in the crusts of x-ray pulsars to model magnetic field evolution \cite{Yakovlev1980,Brown1998,Melatos2001}. This is particularly important as gravitational wave observations, now possible with LIGO and other detectors, provide a window into crust magnetic fields that may differ substantially from surface magnetic fields linked to electromagnetic observables \cite{Mastrano2012}.  

The impact of reaction rate uncertainties in ($\alpha$,p) reactions along the $\alpha$p-process, including  $^{34}$Ar($\alpha$,p)$^{37}$K, on x-ray burst models has been demonstrated in various sensitivity studies \cite{Fis04,Fis08,Par08,Cyb14}. Indirect experimental studies of ($\alpha$,p) reactions along the $\alpha$p-process path have provided indications that rates may differ by orders of magnitude from commonly used Hauser-Feshbach statistical model predictions. Reaction rate estimates based on compound nucleus levels identified in (p,t) transfer reactions are orders of magnitude lower than statistical model predictions for $^{22}$Mg($\alpha$,p)$^{25}$Al \cite{Matic2011}, $^{26}$Si($\alpha$,p)$^{27}$P \cite{Almaraz2012}, $^{30}$S($\alpha$,p)$^{33}$Cl \cite{Long2018}, and $^{34}$Ar($\alpha$,p)$^{37}$K \cite{Long17} (up to a factor of 800 for $^{34}$Ar($\alpha$,p) specifically), questioning the applicability of the statistical model approach and possibly indicating the importance of individual resonances and $\alpha$-clustering effects. A recent measurement of $^{37}$K proton scattering \cite{Lauer17} and tentative evidence from the time-inverse $^{37}$K(p,$\alpha$)$^{34}$Ar reaction (Fig. 3.7 of Ref. \cite{Lauer17}) seem to confirm this picture for $^{34}$Ar($\alpha$,p)$^{37}$K. However, indirect studies of compound nucleus levels may provide incomplete information due to the potential selectivity of the reaction mechanism, and measurements of the time-inverse process only probe the ($\alpha$,p) reaction channel to the ground state (p0). Direct measurements are therefore needed.

At lower masses, a recent direct measurement of the $\alpha$p-process reaction $^{22}$Mg($\alpha$,p)$^{25}$Al \cite{Ran20} found that the cross section is lower than statistical model predictions by factors of 8-10 over the relevant energy range. In contrast, above the $^{34}$Ar waiting point, alpha-induced reactions in the A$\sim$40--50 mass range \cite{Mohr15} show a reasonable ($\sim$2-3x) agreement with Hauser-Feshbach predictions. Here we provide results of the first direct measurement constraints for the $^{34}$Ar($\alpha$,p)$^{37}$K and $^{34}$Cl($\alpha$,p)$^{37}$Ar reactions to address whether the predicted large uncertainties in the applicability of Hauser-Feshbach model rates indeed affects this important mass region of neutron-deficient nuclei.

To constrain the $^{34}$Ar($\alpha$,p)$^{37}$K cross section directly, a spectroscopic measurement was undertaken with the Jet Experiments in Nuclear Structure and Astrophysics (JENSA) gas jet target \cite{Chi14,Schmidt2018,CAARI,NIC}, the most dense helium jet target for RIB studies in the world. The system was operated with research grade helium to provide an average areal density of $\sim 6 \times 10^{18}$ atoms/cm$^2$, and surrounded by the high-segmentation, high-resolution SuperORRUBA \cite{Bardayan2013} and SIDAR \cite{SIDAR} silicon detector arrays for observing the light reaction ejectiles. A position-sensitive ionization chamber (IC) \cite{Lai2018} downstream was used to detect both the heavy recoil and unreacted incident beam.

The $^{34}$Ar beam was produced through fragmentation of a 150 MeV/u, $\sim$60 pnA $^{36}$Ar primary beam from the National Superconducting Cyclotron Laboratory on a 1089 mg/cm$^2$ $^9$Be production target. The secondary beam was separated using the A1900 fragment separator \cite{Mor03}, stopped in a gas cell \cite{Coo14}, charge-bred to 15+ \cite{Lap18}, and reaccelerated using the ReA3 linac. Two energies near the astrophysically-relevant region were measured: $E_{cm} = $ 5.909 $\pm$ 0.079 and 5.607 $\pm$ 0.077 MeV at the center of the target. Total beam intensities on target ranged between $\sim$(2-8)$\times$10$^3$ pps.


While the A1900 and gas cell were able to purify the $^{34}$Ar from the other fragmentation products, the two isobaric decay products, $^{34}$Cl and $^{34}$S, were present in the final beam on target. Stable $^{34}$S was anticipated as $^{nat}$S is present in the system; however, the substantial buildup of $^{34}$Cl from the decay of $^{34}$Ar in the beam was not anticipated. Due to the very similar ($\Delta \sim 100$ keV) Q-values for the ($\alpha$,p) reactions on $^{34}$Cl and $^{34}$Ar, these two contributions could not be separated in the experiment. The isobars in the unreacted beam were separable in the ionization chamber, but due to the steep kinematics of the ($\alpha$,p) reactions, many of the heavy (A=37) recoils stopped in the IC prior to the energy loss sections, preventing reliable separation from standard energy loss techniques. The beam composition was typically in an equilibrium population of approximately 60\% $^{34}$Ar, 30\% $^{34}$Cl, and 10\% $^{34}$S, with only small variations observed over the course of the measurement.

An ($\alpha$,p) reaction event was identified by requiring that a proton signal was observed in the silicon array in timing coincidence with a heavy recoil in the ionization chamber. Protons were identified using standard $\Delta$E-E energy loss techniques in the forward angles of the array. Backward of 90$^{\circ}$ only protons are kinematically allowed. Silicon events without $\Delta$E-E in the forward-angle detectors, and with detected energies of less than 500 keV, were rejected as noise. Events with a recoil angle $<$0.6$^{\circ}$ as determined from the position measurement in the IC were excluded as well to avoid random coincidences with the unreacted beam. This did not result in a significant loss of good events as protons emitted in coincidence with actual recoils at such small angles would have mostly been too low in energy to detect.

\begin{figure} 
\includegraphics[scale=0.43,angle=0]{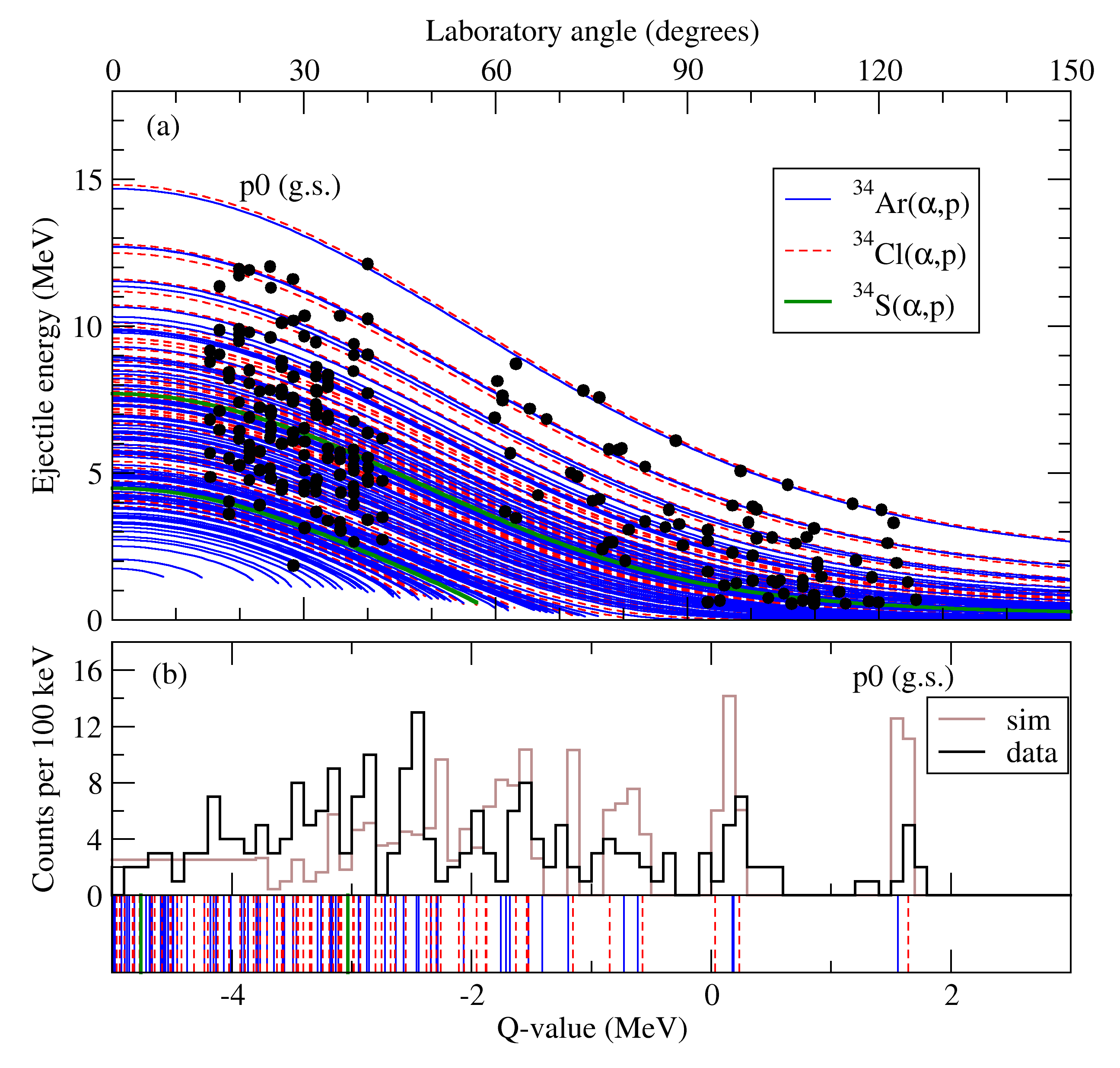}
\caption{\label{kin} (Color online) Calculated two-body kinematics for the ($\alpha$,p) reaction on the three beam constituents ($^{34}$Ar - blue solid; $^{34}$Cl - red dash; $^{34}$S - green bold solid), with the events that met the selection criteria from the current experiment overlaid in black, for the lower beam energy. The upper panel (a) shows the energy versus angle dependence for the identified proton events (black circles) compared with the calculated kinematics curves. The lower panel (b) shows those same events projected into a $^{34}$Ar($\alpha$,p)$^{37}$K Q-value spectrum (black histogram), again compared against the locations of the known levels in $^{37}$K,Ar,Cl. While the resolution is not sufficient to differentiate between excited states in $^{37}$K and $^{37}$Ar, the (combined) p0 channel is clearly delineated. The ground state of the contaminant $^{34}$S($\alpha$,p)$^{37}$Cl reaction falls at a reconstructed Q-value of about -3 MeV. In brown is a simulation of the reconstructed Q-value using the partial cross sections (p0,p1...pN, corresponding to reactions proceeding through the final nucleus ground state, first excited state...Nth excited state, respectively) as predicted by TALYS but with the total cross section normalized to reproduce the number of events seen in the measurement, for comparison.}
\end{figure}

The data are compared with the known reaction kinematics \cite{Wang2017} in Figure \ref{kin} for the lower beam energy as an example. Only two states in the final $^{37}$Cl nucleus, populated via $^{34}$S($\alpha$,p), are kinematically accessible. However, many final states are accessible in the $^{34}$Cl($\alpha$,p)$^{37}$Ar and $^{34}$Ar($\alpha$,p)$^{37}$K reactions. 

The detection efficiencies were obtained from a Monte Carlo simulation of the full setup. The efficiency was determined as a function of excitation energy and angle, and took into account the various gates applied to the data, such as the energy (electronics) thresholds, timing cuts, and recoil angle. The polar and azimuthal angles of the emitted protons were chosen at random from an isotropic distribution in the center of mass, and the weighting for the final state excitation energies was determined using the relative strengths from Hauser-Feshbach calculations using the default parameters in TALYS-1.8 \cite{TALYS}. Sensitivity in the detection efficiency to the details of this excitation energy distribution were found to be $\leq$5\%, by comparing the simulated efficiency for the full predicted distribution to the simulated ground state (p0) efficiency. The detection efficiency for most of the range of protons covered in this measurement was essentially flat, and the impact of the TALYS-predicted partial strengths on the final cross sections was negligible. As a cross-check, the measured rate of scattered alphas was compared to calculations of Rutherford scattering, and found to agree within the final adopted uncertainties.

\begin{figure}
\includegraphics[scale=0.37,angle=0]{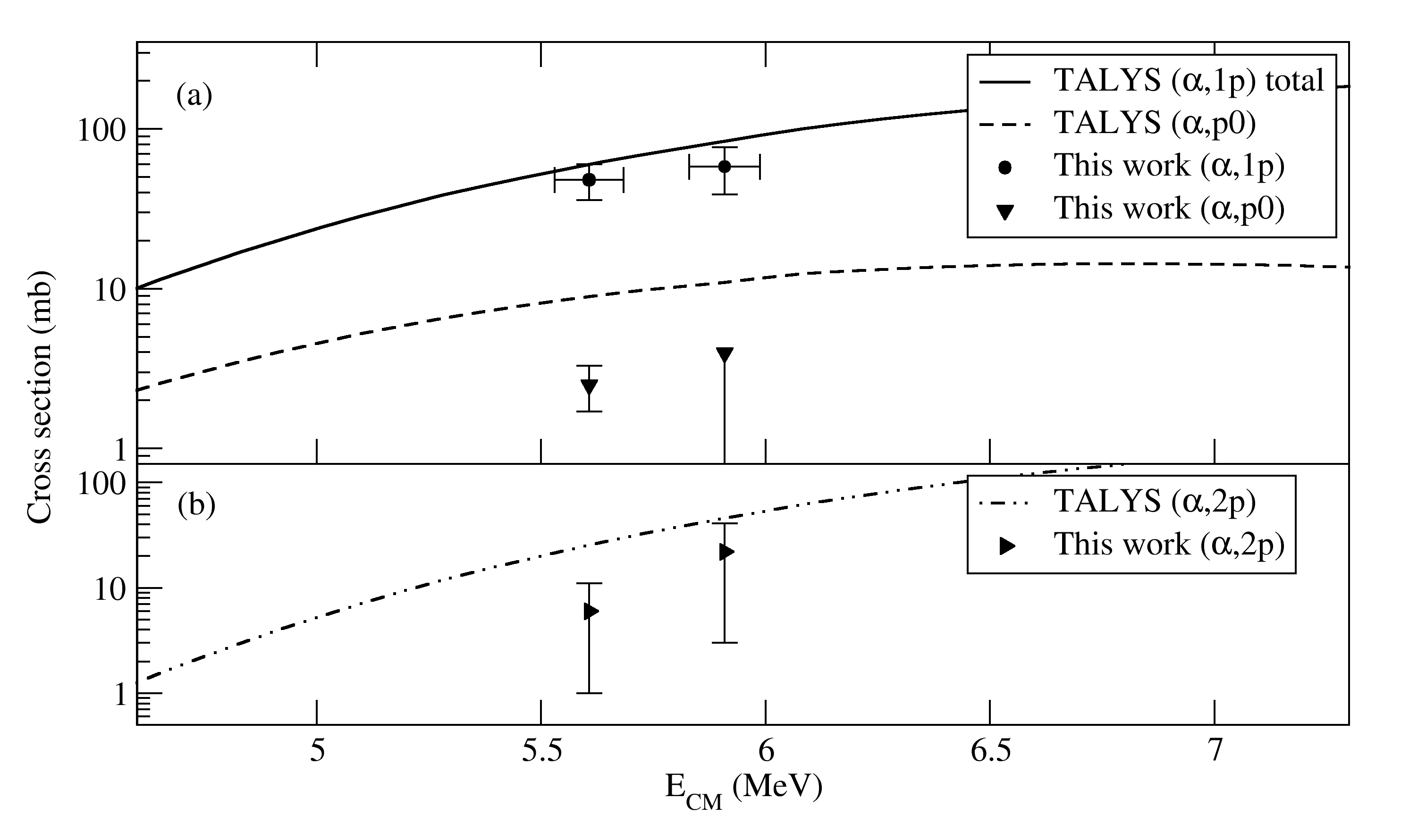}
\caption{\label{apxs} Combined experimental cross sections derived in the current work, compared with Hauser-Feshbach statistical calculations using the default parameters in TALYS. In the upper panel (a), the TALYS calculations for the combined $^{34}$Ar,Cl($\alpha$,1p)$^{37}$K,Ar cross section is shown in black solid, and the p0 contribution in black dashed. Similarly, the $^{34}$Ar($\alpha$,2p)$^{36}$Ar reaction is shown in the lower panel (b). The horizontal error bars represent the uncertainty on the beam energy; the target thickness was about 200 keV in the center of mass.}
\end{figure}

Reaction yields were determined based on the total number of detected ($\alpha$,p) events. ($\alpha$,1p) and ($\alpha$,2p) yields were extracted using the measured multiplicity distribution and simulated proton efficiencies. For $^{34}$Ar, both the ($\alpha$,1p) and the ($\alpha$,2p) channels are energetically possible. For $^{34}$Cl and $^{34}$S only the ($\alpha$,1p) channel is open. The ($\alpha$,2p) yield was therefore exclusively attributed to the $^{34}$Ar($\alpha$,2p)$^{36}$Ar reaction. The $^{34}$S($\alpha$,p) reaction yield to the ground state was subtracted from the ($\alpha$,1p) yield using the known cross section at each beam energy \cite{SCOTT1993363}; the yield to the first excited state was taken to be negligible at these energies. However, the cross section of the $^{34}$Cl($\alpha$,p) reaction is not known, and the overlap of excited states in the $^{37}$Ar and $^{37}$K final nuclei for these two ($\alpha$,p) reactions (Fig. \ref{kin}) meant such a subtraction was not possible. Hence the final cross sections reported here (Figure \ref{apxs} and Table \ref{apresult}) are for the combined $^{34}$Ar,Cl($\alpha$,1p)$^{37}$K,Ar reactions.

Theoretical cross sections for the $^{34}$Ar($\alpha$,1p/2p) and $^{34}$Cl($\alpha$,1p)$^{37}$Ar reactions were calculated with TALYS-1.8 \cite{TALYS}, adopting the default level density and alpha-nucleus optical models. The calculated $^{34}$Ar($\alpha$,1p) and $^{34}$Cl($\alpha$,1p) cross sections were combined in a weighted sum, taking into account the relative intensities of the beam constituents (for the higher beam energy, this ratio was 32.3\% to 67.7\% $^{34}$Cl to $^{34}$Ar, respectively; for the lower energy, 33.5\% to 66.5\%). The theoretical predictions are compared with experimental results in Fig. \ref{apxs}. The measured $^{34}$Ar,Cl($\alpha$,1p) cross sections agree remarkably well with Hauser-Feshbach predictions. The $^{34}$Ar($\alpha$,2p)$^{36}$Ar cross section is somewhat overpredicted but, even with the default, non-tuned input parameters, agrees with the data within the factor of 2-3 often considered the expected precision of the statistical model approach \cite{Mohr15}, lending additional weight to the validity of the Hauser-Feshbach predictions. Most importantly, we can exclude statistical models overpredicting the cross section by several orders of magnitude, as was inferred from indirect measurements in this mass region \cite{Long17,Lauer17}, at the energies probed in this experiment.

The high-resolution proton spectroscopy enabled by JENSA allowed determination of both the total cross section and the partial cross section to the final ground state p0 (Fig. \ref{kin}), as well as the population distribution of excited final states. The ratio of the ground state transition to the total cross section informs the time-inverse reaction studies -- in this case, $^{37}$K(p,$\alpha$)$^{34}$Ar \cite{DeibelNICXI,DeibelNICXII} -- which use the detailed balance principle between the (p0,$\alpha$0) and ($\alpha$0,p0) ground state channels but rely on statistical model corrections to account for excited states. Our data can be directly compared to the Hauser-Feshbach predicted strengths to each final partial state (Fig. \ref{kin}b). We experimentally determine a combined p0 branch of 5$\pm$2\% at the lower beam energy, compared to a combined 10.4\% predicted by TALYS (the predicted partial strengths as a function of excitation energy for the two beam constituents follow the same trend). This shows that time-inverse studies probing the ground state component alone do not necessarily provide a reliable test for reaction rate predictions of the total rate. Figure \ref{kin}b also compares the predicted population distribution of excited final states p0...pN with the experimental data. There are discrepancies with the statistical models predicting a stronger population of lower excitation energies compared with the experimental data, in line with the result for p0. Despite this, the overall cross section agrees well.

\begin{table}
\caption{\label{apresult}Combined cross sections (mb) from this work and from default TALYS Hauser-Feshbach predictions. The range of TALYS predictions, dominated by the choice of alpha-nucleus optical model, encompasses roughly 2-3x around these values.}
\begin{ruledtabular}
\begin{tabular}{l|cc|cc}
 &  \multicolumn{2}{c}{5.6~MeV} & \multicolumn{2}{c}{5.9 MeV} \\
 & This Work & TALYS & This Work & TALYS \\
\hline
$^{34}$Ar,Cl($\alpha$,xp) & $54\pm13$ & 85.5 & $80\pm27$ & 129.4\\
$^{34}$Ar,Cl($\alpha$,1p) & $48\pm12$ & 60.0 & $58\pm19$ & 83.7 \\
$^{34}$Ar($\alpha$,2p) & $6\pm5$ & 25.5 & $22\pm19$ & 45.7 \\
$^{34}$Ar,Cl($\alpha$,p0) & $2.5\pm0.8$ & 8.9 & $\leq2.7$ & 10.9\\
\hline
\end{tabular}
\end{ruledtabular}
\end{table}

\begin{figure} 
\includegraphics[scale=0.45,angle=0]{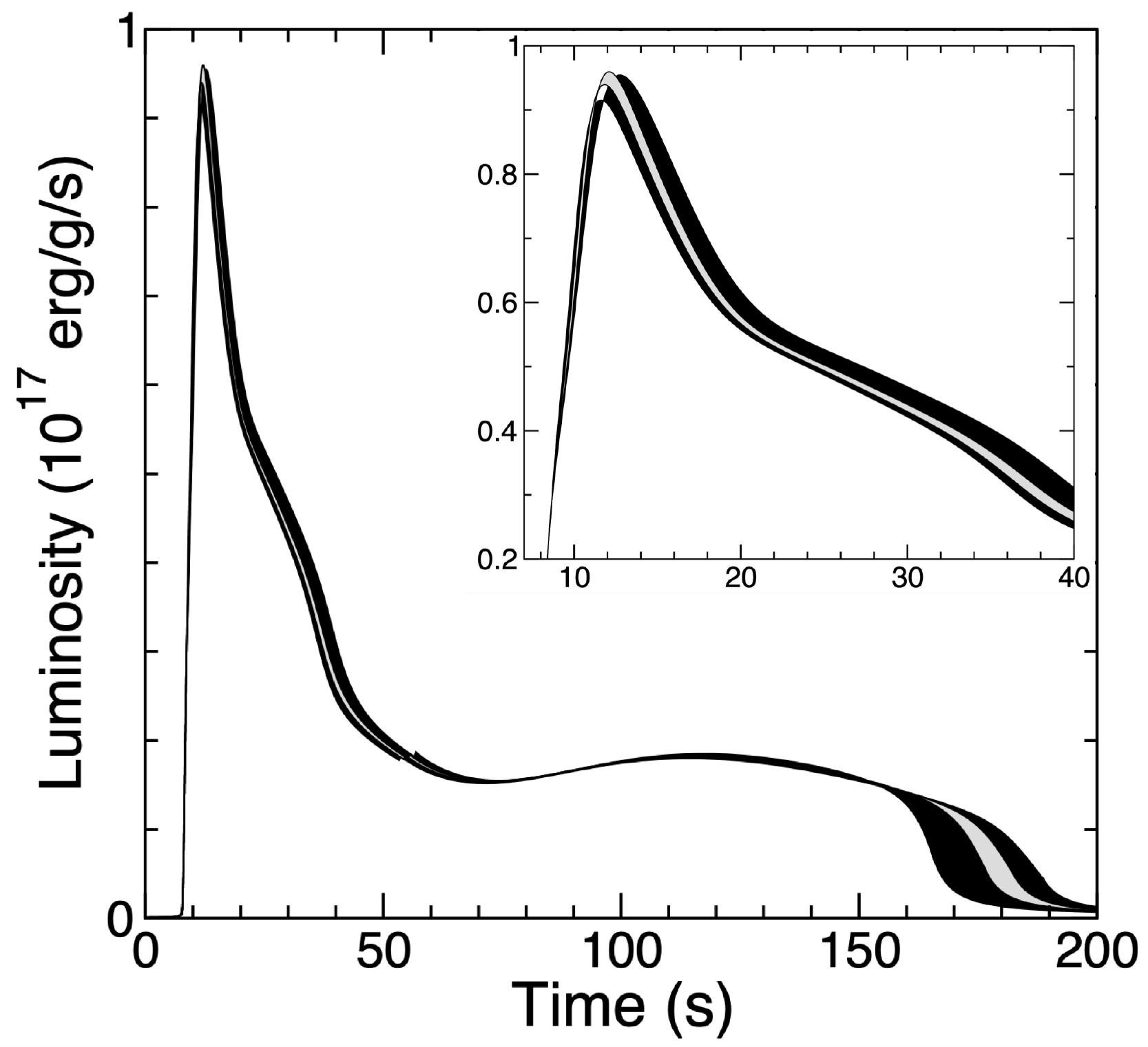}
\caption{\label{fig:xrb} X-ray burst light curve in luminosity per gram of accreted matter predicted by one-zone model \cite{Schatz2001} for a $^{34}$Ar($\alpha$,p)$^{37}$K reaction rate reduced by a factor of 100 and multiplied by a factor of 10 (shaded black), and for the same reaction rate reduced and multiplied by a factor of 2 (shaded gray). The inset zooms in on the burst peak. The small kink around 55 seconds stems from a switch between tabulated data and an analytic approximation used in the opacity calculation.}
\end{figure}

To estimate the astrophysical impact of these results, we compare astrophysical simulations obtained with variations of the $^{34}$Ar($\alpha$,p)$^{37}$K rate by a factor of 10 up and a factor of 100 down (reflecting previously estimated uncertainties) and calculations for a variation of $\times$2, in line with the confirmation in this work that typical statistical model uncertainties are applicable. For x-ray bursts we use a one-zone model of a high accretion rate and low metallicity burst that leads to an extended rp-process \cite{Schatz2001}. The results are shown in Figure~\ref{fig:xrb}. With the previous reaction rate uncertainties, the peak luminosity, initial burst width, and length of the burst tail are all affected significantly. In particular, a lower rate leads to a shorter burst tail due to the significantly more rapid hydrogen fuel consumption. This is a consequence of a smaller rate producing fewer heavy seed nuclei for hydrogen capture, thus decreasing the hydrogen-to-seed ratio resulting in a shortened rp-process. A shorter rp-process includes fewer slow $\beta$-decays and consumes the hydrogen faster. Within the $\times$2 uncertainties adopted based on this work, there is no significant contribution from the $^{34}$Ar($\alpha$,p)$^{37}$K reaction rate on x-ray burst light curve model predictions.

\begin{figure} 
\includegraphics[scale=0.35,angle=0]{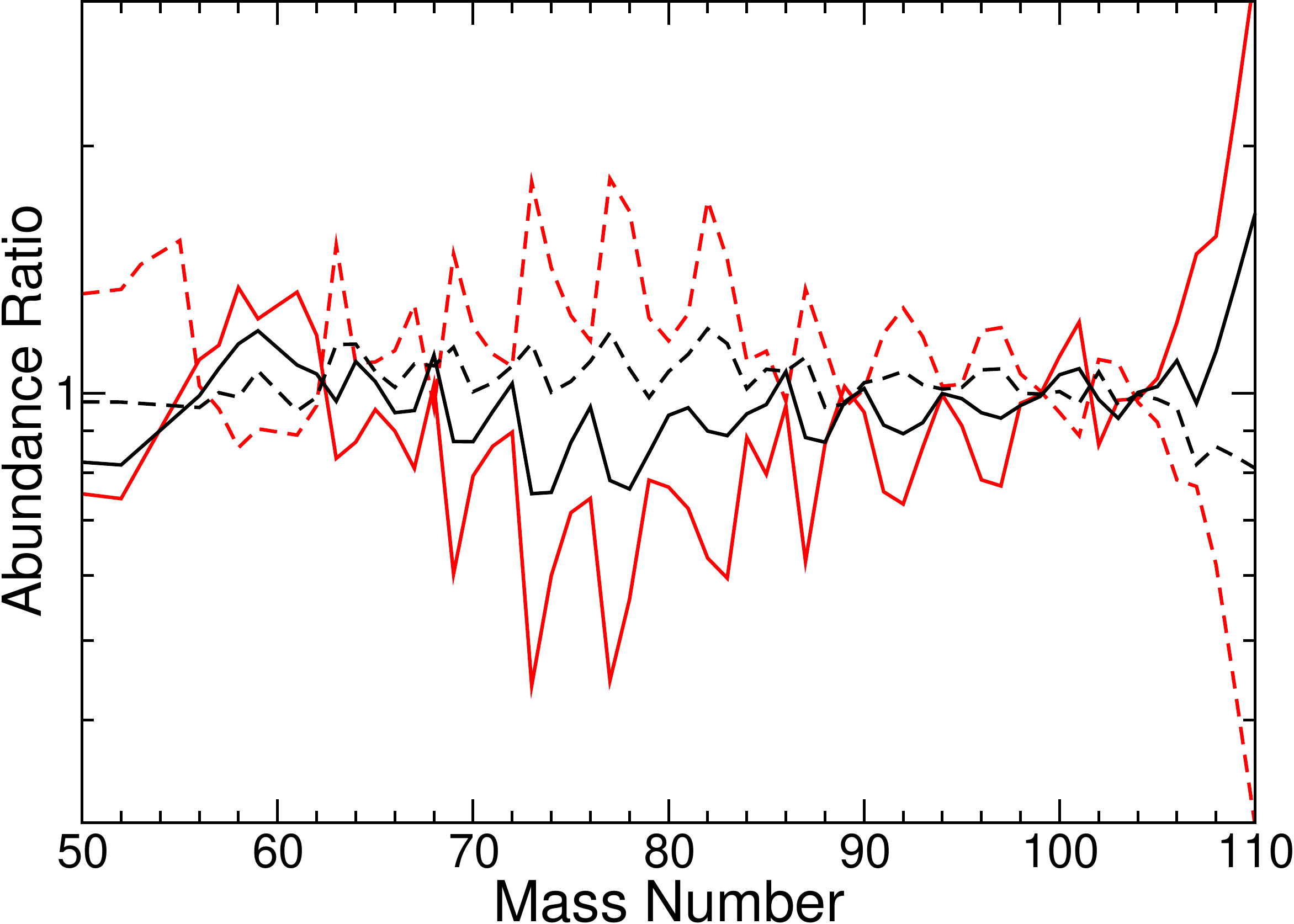}
\caption{\label{fig:ashes} Impact of variations of the $^{34}$Ar($\alpha$,p)$^{37}$K reaction rate on the final composition of the nuclear ashes in steady state burning at a local accretion rate of 40~$\dot{m}_{\rm Edd}$ \cite{Schatz1999}. Shown are the ratios of final abundances, summed by mass number,  as functions of mass number relative to the baseline calculation for a variation of the reaction rate by a factor of 10 (solid red), 0.01 (dashed red), 2 (solid black), and 0.5 (dashed black). Ratios are only displayed for total abundances larger than 10$^{-6}$ mole/g.}
\end{figure}

For steady state nuclear burning, we use the model described in \cite{Schatz1999} with a base flux from the neutron star crust of 0.5~MeV/nucleon. For a factor of 100 variation of the $^{34}$Ar($\alpha$,p)$^{37}$K reaction rate, we find significant differences in the composition of the nuclear ashes for local accretion rates of $\dot{m}=20$~$\dot{m}_{\rm Edd}$ and higher (in units of the Eddington accretion rate $\dot{m}_{\rm Edd}=8.8 \times 10^4$~g s$^{-1}$ cm$^{-2}$). At such accretion rates, the hydrogen burning temperature is sufficiently high for the $\alpha$p-process to reach the Ar region. For example, for 40~$\dot{m}_{\rm Edd}$ , abundance variations for an increase of the rate by a factor of ten and decrease by a factor of 100 can span an order of magnitude (Fig.~\ref{fig:ashes}). For a $\times$2 rate uncertainty, abundance variations are less than a factor of 1.7, a significant improvement on the experimental constraint.

In summary, the first direct measurement of the $^{34}$Ar($\alpha$,2p)$^{36}$Ar and combined $^{34}$Ar,Cl($\alpha$,1p)$^{37}$K,Ar cross sections demonstrates that, in contrast to inferences from indirect reaction data, the statistical model is likely applicable in this mass region for predictions of astrophysical ($\alpha$,p) reaction rates on neutron-deficient nuclei. The experimentally-constrained cross sections are within the expected uncertainty factors of $\sim$2-3 for the statistical model approach, and in line with both findings for reactions with heavier nuclei and more detailed theoretical approaches \cite{Brune2018}. The confirmation of the applicability of statistical approaches to calculate ($\alpha$,p) reaction rates in the A$\sim$30 mass region, in contrast to earlier indirect results, removes a major uncertainty in astrophysical models of accreting neutron stars. As recent direct measurements indicate a factor of 8 overestimation of the $^{22}$Mg($\alpha$,p) reaction rate by statistical models, more measurements in the $A=22-34$ mass range, for example of $^{26}$Si($\alpha$,p) or searches for alpha cluster states, would be important to experimentally determine whether there is a minimum mass range for reliable application of statistical models. Finally, while our measurements are performed at energies where the predictions from indirect reaction data can be tested, they are higher than the relevant astrophysical energies of $E_{\rm CM}=2-3.9$~MeV. Future measurements at lower energies are needed to confirm statistical models can be used to extrapolate cross sections to the full astrophysical energy range. 

\begin{acknowledgements}The authors would like to thank all those who made contributions to JENSA and E15232. This material is based upon work supported by the U.S. DOE, Office of Science, Office of Nuclear Physics under contracts DE-AC05-00OR22725 (ORNL) and DE-FG02-93ER40789 (Colorado School of Mines), and by the National Science Foundation under Award Numbers PHY-1430152 (JINA Center for the Evolution of the Elements), PHY-1565546 (NSCL), PHY-1913554 (MSU), and PHY-2011890 (Notre Dame). Research sponsored by the Laboratory Directed Research and Development Program of Oak Ridge National Laboratory, managed by UT-Battelle, LLC, for the U.S. Department of Energy. This work also supported partially by the National Research Foundation of Korea (NRF) grants funded by the Korea government (Grant No. 2020R1A2C1005981).
\end{acknowledgements}



\end{document}